# Primitive ontology and quantum state in the GRW matter density theory

Matthias Egg and Michael Esfeld
University of Lausanne, Department of Philosophy
Matthias.Egg@unil.ch, Michael-Andreas.Esfeld@unil.ch

Abstract

The paper explains in what sense the GRW matter density theory (GRWm) is a primitive ontology theory of quantum mechanics and why, thus conceived, the standard objections against the GRW formalism do not apply to GRWm. We consider the different options for conceiving the quantum state in GRWm and argue that dispositionalism is the most attractive one.

*1.* Introduction

A popular summary of the measurement problem in quantum mechanics (QM) is contained in the following quote from Bell (1987, p. 201): "Either the wavefunction, as given by the Schrödinger equation, is not everything, or it is not right". This suggests two principled ways to solve the problem, either by adding something to the Schrödinger equation or by modifying it. (A third option, which we will not consider here, would be to revise our beliefs about the uniqueness of measurement outcomes. This is the Everettian way of avoiding Bell's dilemma, leading to a kind of "many worlds" interpretation of QM.) The most well known elaboration of the second solution is the GRW theory (Ghirardi, Rimini and Weber 1986).[1] This theory features a modified Schrödinger evolution assigning to quantum systems a probability to undergo spontaneous localizations ("collapses") at random times. This probability is very small for microscopic systems, but it approaches 1 for macroscopic systems. The GRW theory therefore promises to ensure what the unmodified Schrödinger evolution cannot achieve, namely the unique localization of macroscopic bodies and, more specifically, unique measurement outcomes.

However, it has become increasingly clear in the last two decades that modifying the Schrödinger equation is, by itself, not sufficient to solve the measurement problem. The reason is that the GRW equation still is an equation that describes the temporal development of the wave-function in configuration space, by contrast to an equation that describes the temporal development of matter in physical space. In today's literature, the distribution of matter in physical space is known as the "primitive ontology" (that term goes back to Dürr, Goldstein and Zanghì 2013, section 2.2, originally published 1992). It is primitive in the sense that it cannot be inferred from the formalism of textbook QM, but has to be put in as the referent of that formalism. The motivation for doing so is to obtain an ontology that can account for the existence of measurement outcomes – and, in general, the existence of the macrophysical objects with which we are familiar before doing science. Hence, what is introduced as the primitive ontology has to be such that it can constitute measurement outcomes and localized macrophysical objects in general.

---

[1] The prominence of the GRW theory may be somewhat surprising, because the physics literature features collapse theories which are better developed and generally regarded as more realistic, such as the *Continuous Spontaneous Localization* (CSL) theory. These latter theories are, however, mathematically more demanding, whereas they do not seem to differ significantly from the GRW theory on the conceptual and ontological level. For this reason, the philosophical debate has tended to focus on GRW instead of CSL; we will discuss the justification for this move in section 3. For a comprehensive review of different collapse models, see Bassi and Ghirardi (2003).

One way to do this, taking seriously the idea that a quantum system such as an electron does not usually have a determinate value of position, is to say that such systems are smeared out in physical space. What the GRW dynamics then achieves is to describe how this smeared out position develops such as to be concentrated around a point. This is indeed the reading of the physical significance of the GRW dynamics that Ghirardi himself puts forward (see Ghirardi, Grassi and Benatti 1995, sections 3-4): an electron when it does not have a determinate position is literally smeared out in physical space, constituting thus a matter field (in the sense of a continuous distribution of stuff in physical space). Generalizing this idea, the proposal hence is that the wave-function in configuration space represents the density of matter in physical space. This proposal is known as the GRW mass or matter density theory (GRWm). We will introduce it in more detail in section 3.

Apart from GRWm, there are two other elaborate proposals for a primitive ontology of QM. The de Broglie-Bohm theory, going back to de Broglie (1928) and Bohm (1952) and known today as Bohmian mechanics (BM) (see Dürr, Goldstein and Zanghì 2013) is the oldest of them. BM endorses particles as the primitive ontology, maintaining that there is at any time one actual configuration of particles localized in three-dimensional space, with the particles moving on continuous trajectories in space. BM therefore needs two laws:

$$\frac{dQ}{dt} = \frac{\hbar}{m_k} \Im \frac{\psi^* \nabla_k \psi}{\psi^* \psi}(Q) \tag{1}$$

$$i\hbar \frac{\partial \psi}{\partial t} = H\psi \tag{2}$$

The guiding equation (1) fixes the temporal development of the position $Q = (Q_1(t), ... , Q_N(t))$ of the particles, and the Schrödinger equation (2) determines the temporal development of the wave-function. Thus, the role of the wave-function, developing according to the Schrödinger equation, is to determine the velocity of each particle at any time $t$ given the position of all the particles at $t$.

Furthermore, there is another proposal for a primitive ontology of QM that uses the GRW dynamics: Bell (1987, ch. 22) suggests that whenever there is a spontaneous localization of the wave-function in configuration space, that development of the wave-function in configuration space represents an event occurring at a point in physical space. These point-events are today known as flashes; that term was introduced by Tumulka (2006a, p. 826). According to the GRW flash theory (GRWf), the flashes are all there is in space-time.

These theories hence put forward different proposals about the nature of matter, which cover the main metaphysical conceptions of matter – particles or atoms in BM, a field in the sense of gunk in GRWm, and single events in GRWf. Nonetheless, their structure is the same: they consist in a proposal for a primitive ontology of matter distributed in physical space and a law for its temporal development (see Allori et al. 2008). Consequently, the role of the quantum formalism in these theories is not to provide the ontology of the theory. Its role is this one: given an ontology in the guise of the primitive ontology, the role of the quantum formalism – in particular the wave-function and the law in which it figures – is to describe the temporal development of the primitive ontology, that is, the distribution of matter localized in three-dimensional space.

Against this background, it is well-motivated to take the view that the primitive ontology is primitive yet in another sense: it is primitive not only in the methodological sense that it has to be put in as the referent of the quantum formalism that one adopts, but also in the ontological sense that it consists in primitive stuff. Thus, Allori et al. (2014, pp. 331-332) say: "Moreover, the matter that we postulate in GRWm and whose density is given by the *m* function does not *ipso facto* have any such properties as mass or charge; it can only assume various levels of density." Consequently, the ontology of GRWm is not one of different kinds of particles, with the individual particles being smeared out in physical space (unless their wave-function spontaneously localizes in configuration space) and thus building up overlapping matter density fields. There is just one matter density field that fills all space and that is *materia prima*, primitive stuff, being more dense in some regions of space and less dense in other regions.

Mass and charge then enter into the picture as dynamical parameters playing a certain role for the temporal development of the distribution of the primitive stuff in space. The same goes for the wave-function, which represents the quantum state of a system: the quantum state is not a state in the sense of something static, but a dynamical variable performing a certain function for the temporal development of the distribution of the primitive stuff in space. However, whereas mass and charge can be considered as properties that the primitive stuff has at points of space, the quantum state cannot be treated as such a property. The reason is quantum entanglement. Strictly speaking, there is only one entangled, universal quantum state for the whole configuration of the stuff in physical space, which is represented by the universal wave-function. Hence, the question is what the ontological status of this wave-function – more precisely, the quantum state as represented by the wave-function – is in these theories.

This question has mainly been investigated with respect to BM so far. The purpose of this paper is to extend this investigation to GRWm. As we will see, there are considerable differences between BM and GRWm as regards the options for conceiving the quantum state in these theories. Accordingly, we first review the options for an ontology of the quantum state in BM (section 2) and then discuss the basic elements of GRWm, showing how this theory overcomes the objections traditionally raised against the GRW approach (section 3). In section 4, we analyse in a rather general fashion the possible relations of ontological priority between the primitive ontology and the quantum state. Building upon the preparatory work performed in sections 2-4, section 5 then assesses the different options for understanding the quantum state in the context of GRWm.

2.     *The quantum state in Bohmian mechanics: a spectrum of options*

In the context of BM, Belot (2012) has outlined three main options for understanding the quantum state of the Bohmian particle configuration of the universe as represented by the universal wave-function $\psi$: either as an object, or as a law, or as a property. He further subdivides the first option into two proposals depending on the kind of object that one takes the quantum state to be. Its mathematical representation clearly suggests that it is some kind of field, but it is equally clear that it cannot be an ordinary field on three-dimensional space or four-dimensional space-time, at least not if one considers $N$-particle systems with $N > 1$. One then has to regard the quantum state either as a field on $3N$-dimensional configuration space or as what Belot calls a "multi-field", an object assigning properties to $N$-tuples of points in space.

The second of Belot's three options admits of even more interesting subdivisions, corresponding to the different conceptions of laws of nature put forward in metaphysics. Esfeld et al. (2014) discuss the two main contenders, namely *Humeanism* and *dispositionalism*, and they briefly mention *primitivism* as a further option. On the Humean view, the quantum state is not part of the fundamental ontology of BM. Fundamentally, particles and their trajectories in space and time are all there is. Everything else supervenes on the particle positions throughout the whole of space-time, which constitute the Humean mosaic of local matters of particular fact on Bohmian Humeanism. That is to say, the ontological status of $\psi$ is exhausted by the role that it plays in the Humean best system, that is, the system that achieves the best balance between being simple and being informative in describing the particle positions throughout the whole of space-time (see Miller 2014, Esfeld 2014 and Callender 2014; cf. already Dickson 2000).

By contrast, dispositionalism takes $\psi$ to refer to an additional element in the Bohmian ontology, namely a holistic property of the particle configuration as a whole that is a modal property, that is, a disposition or power of the particles to move in a certain way (or to stay at rest). That disposition or power manifests itself in the velocity of the particles. This is the way in which dispositionalism spells out the idea that there is something in BM that guides or pilots the motion of the particles (see Esfeld et al. 2014, sections 4-5). On this view, $\psi$'s nomological role is grounded in the physical property to which $\psi$ refers, as dispositionalism in general takes the laws of nature to be fixed by the dispositions or powers that there are in the world (see Bird 2007 for the general view). It is therefore more appropriate to regard this view as spelling out Belot's third option ($\psi$ as property) than as a subtype of the second option ($\psi$ as law).

Primitivism about laws shares with dispositionalism the anti-Humean intuition that there is something in the ontology that governs the particles' behaviour, but it denies the need for laws to be grounded in anything. Instead, over and above the primitive ontology, there is in any possible world a fact instantiated that certain dynamical laws hold in the world in question. That fact determines – or at least puts a constraint on – how the initial configuration of the primitive ontology develops in time (see notably Maudlin 2007 for primitivism about laws in philosophy of physics). The distinction between a nomological and an ontological understanding of $\psi$ collapses in this case, since, if $\psi$ refers to some element of a law, it *ipso facto* refers to an element of the ontology.

We can classify these different views of the quantum state according to the extent to which they grant this state an autonomous status in the ontology of physics. On the one end of the spectrum, we find the view of $\psi$ as an object (a field or multi-field), which attributes to the quantum state the most autonomous status, since it treats it as a physical object of its own. Next comes the view of $\psi$ as nomological in the framework of primitivism about laws of nature, since in this framework, the view of $\psi$ as nomological amounts to admitting an autonomous fact of a certain law being instantiated in the domain of physics. Then comes the view of $\psi$ being a disposition of the configuration of the primitive ontology. In this case, $\psi$ is a property of the primitive ontology and thus instantiated by the primitive ontology; but it is not determined by the primitive ontology: the same initial configuration of the primitive ontology can go with different quantum states, which manifest themselves in differences in the temporal evolution of the initial configuration. By contrast, on the nomological view of $\psi$ as spelled out in the framework of Humeanism about laws of nature, the quantum state is determined by the primitive ontology in supervening on it; but it supervenes only on the distribution of the elements of the primitive ontology throughout the whole of space and time (the Humean mosaic). Note that all these views are committed to realism about the quantum state. Even on the Humean view, the quantum state exists, but it does not have any autonomous status in supervening on the primitive ontology as a whole. It hence is not fundamental. Nonetheless, non-fundamental entities, which supervene on or are reducible to other entities, are real (Miller 2014, section 5, stresses this point).

## 3.    *The matter density ontology*

Allori et al. (2008, p. 359) give the following concise characterization of GRWm:

> We have a variable *m(x,t)* for every point $x \in \mathbb{R}^3$ in space and every time *t*, defined by
>
> $$m(x,t) = \sum_{i=1}^{N} m_i \int_{\mathbb{R}^{3N}} dq_1 \cdots dq_N \delta(q_i - x) |\psi(q_1,\ldots,q_N,t)|^2 \quad . \quad [(3)]$$
>
> In words, one starts with the $|\psi|^2$-distribution in configuration space $\mathbb{R}^{3N}$, then obtains the marginal distribution of the *i*th degree of freedom $q_i \in \mathbb{R}^3$ by integrating out all other variables $q_j$, $j \neq i$, multiplies by the mass associated with $q_i$, and sums over *i*. ... The field $m(\cdot, t)$ is supposed to be understood as the density of matter in space at time *t*. Since these variables are functionals of the wave function $\psi$, they are not 'hidden variables' since, unlike the positions in BM, they need not be specified in addition to the wave function, but rather are determined by it. Nonetheless, they are additional elements of the GRW theory that need to be posited in order to have a complete description of the world in the framework of that theory.

Thanks to its unambiguous description of matter in space and time, GRWm is not hit by the standard objection to the GRW approach, the so-called "problem of tails". More precisely, GRWm offers a straightforward solution to what Wallace (2008, p. 56) calls the *problem of bare tails*. The problem arises from the fact that the GRW theory mathematically implements spontaneous localization by multiplying the wave-function with a Gaussian, such that the collapsed wave-function, although being sharply peaked in a small region of configuration space, does not actually vanish outside that region; it has tails spreading to infinity. In the literature starting with Albert and Loewer (1996) and Lewis (1997), it is therefore objected that the GRW theory does not achieve its aim, namely to describe measurement outcomes in the form of macrophysical objects having a definite position. However, there is nothing indefinite about the positions of objects according to GRWm. It is just that an (extremely small) part of each object's matter is spread out through all of space. But since the overwhelming part of any ordinary object's matter is confined to a reasonably small spatial region, we can perfectly well express this in our (inevitably vague) everyday language by saying that the object is in fact located in that region (see Monton 2004, pp. 418-419, and Tumulka 2011).

A related problem that has received far less attention in the literature, but is actually more serious for GRWm, is the *problem of structured tails* (Wallace 2008, p. 56). Consider a situation in which the pure Schrödinger evolution would lead to a superposition with equal weight of two macroscopically distinct states (say, a live and a dead cat). The GRW dynamics ensures that the two weights do not stay equal, but that one of them (e.g., the one pertaining to the dead cat) approaches unity while the other one becomes extremely small (but not zero). In terms of matter density, we then have a high-density dead cat and a low-density live cat. The problem now is that the low-density cat is just as cat-like (in terms of shape, behaviour etc.) as the high-density cat. Thus, it is unclear on what grounds we could disregard it as unreal. But then it seems that GRWm is actually a many-worlds theory, as Maudlin (2010, p. 138) points out:

> There is one high-density world and many, many low-density worlds, largely uninfluenced by one another, each with a right to be called 'macroscopic' and 'physically real'. One is, sure enough, of much higher density than the rest, but all are equally real, equally (structurally) stable, equally functional. What, then, is the significance of the density? Could we have any reason, right now, to feel assured that *we* are high-density rather than low-density objects?

It is, however, not quite true that the low-density worlds are structurally as stable as the high-density world. Due to the shape of the Gaussian (in particular, its tails), the low-density worlds suffer distortion effects whenever the wave-function collapses (see Wallace 2014). Maudlin (2010, p. 135 note 2) anticipates this reply, but dismisses it "since the choice of the Gaussian was not made with consideration of its tails, save that they be small". But notice that if we follow this line of argument and regard the shape of the Gaussian as a purely mathematical feature devoid of ontological significance, then we have lost the reason to worry about structured tails in the first place. Indeed, as Ghirardi (2002, section 12) emphasizes, "there is nothing in the GRW theory which would make the choice of functions with compact support problematic for the purpose of the localizations". Such a choice would obviously solve the problem of structured tails by completely eliminating the low-density worlds at the moment of wave-function collapse. Of course, due to the spreading of wave packets, tails would immediately reappear (which is why Ghirardi, continuing the sentence just quoted, calls this procedure "totally useless"), but this would merely create a problem of bare tails, which has its solution in GRWm, as mentioned above. Conversely, if we take the tails of the Gaussian ontologically seriously, we should also take the supposed distortion effects seriously, and the fact that we do not observe them should count as evidence that we do indeed inhabit the high-density world and can safely neglect the low-density worlds.[2]

Another problem of the GRW theory (at least in its original form) is that it cannot deal with systems of identical particles, because its collapse mechanism does not conserve the symmetry character of wave-functions describing such systems.[3] This defect is overcome by the Continuous Spontaneous Localisation (CSL) theory (Ghirardi, Pearle and Rimini 1990; see also Gisin 1989), which replaces the instantaneous collapses of GRW by a continuous stochastic evolution of the wave-function. A further advantage of the CSL theory is its ability to describe the complete time evolution of the quantum state by a (stochastic) differential equation, whereas the discontinuous character of the collapses makes such a description impossible for the GRW theory.

---

[2] For further criticism of the claim that the low-density worlds are structurally equivalent to the high-density world, see Albert (forthcoming, chap. 7).

[3] An anonymous referee has pointed out to us that there are variants of the GRW theory which can deal with identical particles. See Tumulka (2006b) for details.

The CSL theory is important for the matter density ontology, not only because the latter was first proposed in the context of the former (Ghirardi, Grassi and Benatti 1995), but also for two more substantial reasons. First, in contrast to the GRW theory, which can just as well accommodate a discrete ontology (GRWf) as a continuous one (GRWm), the CSL theory strongly favours a continuous ontology of the matter density type, as the usual ways to construct a discrete ontology for GRW do not work for CSL (see Bacciagaluppi 2010, section 3). Second, while the flash ontology was for some time thought to possess a decisive advantage over the matter density ontology when it comes to making collapse theories compatible with special relativity (Tumulka 2006a), recent progress in the construction of relativistic CSL-type models (in particular, Bedingham 2011) indicates that the matter density ontology may have equally good prospects for being made relativistic (Bedingham et al. 2014; for a critical assessment of relativistic GRWf, see Esfeld and Gisin 2014).

Nevertheless, the debate about the ontology of spontaneous localization theories has not been much concerned with CSL, but has for the most part focused on the mathematically simpler original GRW theory. This might seem illegitimate, since the ontological picture of a matter density undergoing discontinuous changes (as in GRWm) seems to differ significantly from the CSL picture, according to which the matter density always evolves continuously. But if we remember that both theories were primarily introduced to explain the localization of macroscopic objects, then the difference between them does not seem so significant anymore: in systems consisting of a large number of constituents, the localization process is so fast that the difference between an instantaneous and a gradual process becomes negligible. Put in mathematical terms, it can be shown that the GRW theory approximates the CSL theory in the limiting case of the collapse frequency going to infinity (Nicrosini and Rimini 1990). This reasoning justifies, for many contexts, the choice to look at GRW instead of CSL, and we will follow this methodology for most of our discussion.

Given this somewhat unrealistic character of the GRW theory, we should expect some of its mathematical features not to be directly significant for the ontology. This observation resolves a seeming internal inconsistency of GRWm, which, in spite of its commitment to a field ontology, displays in its mathematical formalism quite an explicit commitment to particles: not only is the matter density (3) defined as a sum of $N$ fields, each one of which is associated with one component of the $N$-"particle" wave-function, but also each GRW collapse

$$\psi \mapsto \frac{L_{\boldsymbol{x}}^i \psi}{\|L_{\boldsymbol{x}}^i \psi\|}, \quad L_{\boldsymbol{x}}^i = K e^{-\alpha(\boldsymbol{q}_i - \boldsymbol{x})^2} \qquad (4)$$

is associated with the position operator $\mathbf{q}_i$ of one of the $N$ "particles". When we turn from the GRW to the CSL formalism, these particle labels disappear in the same way as they disappear in the process known as "second quantization": a quantum theory formulated on a tensor product of $N$ (labelled) one-particle Hilbert spaces is replaced by a theory formulated on a Fock space, in which state descriptions no longer use particle labels, but only occupation numbers.

Therefore, when one follows the above-mentioned methodology and tries to draw ontological conclusions from the GRW formalism, one should regard the particle labels appearing in that formalism (in the first instance, at least) as mere mathematical devices without any ontological significance. What is real is only the total matter density described by $m(x,t)$, not the individual component fields in the sum on the right hand side of (3). As for the collapses, one should remember that the probability of their occurrence is negligibly small unless systems with a large number of constituents are considered. For such systems, due to entanglement among the constituents, it is completely irrelevant *which one* of the constituents is hit by the collapse.

For specific situations, the particle labels can, of course, *acquire* ontological significance, as indeed they should, if the theory is to explain the particle-like behaviour that we observe in many experiments. Such situations arise when subsystems of the universe are suitably isolated from their environment so that they can individually be described by the GRW law (see Goldstein, Tumulka and Zanghì 2012 for details). An example would be what occurs when an experimenter in a laboratory "prepares a one-particle state". The dynamics of the total matter density then is such that one part of it (described by one of the component fields appearing on the right hand side of (3)) behaves autonomously for a certain time. If that part is later allowed to interact with a suitably prepared part of the environment ("the measuring device"), then the GRW dynamics ensures that it manifests itself in the usual way, for example as a black spot on a photographic plate. This is how the GRWm theory accounts for particle phenomenology even though there is nothing particle-like in its fundamental ontology.

### *4.     Primitive ontology and ontological priority*

Before we turn to a detailed discussion of the ontological status of the quantum state in the context of GRWm, let us analyse the ontological relation between the quantum state and the primitive ontology in a more general fashion, as this analysis will impact the assessment to be carried out in the next section. On a rather coarse-grained level, the relative ontological status of the two entities in question can be viewed in either of the following three ways:

(O1)   The primitive ontology is fundamental, the quantum state depends on it.
(O2)   The primitive ontology and the quantum state are both fundamental.
(O3)   The quantum state is fundamental, the primitive ontology depends on it.

If this enumeration is to make sense, we obviously must not equate "primitive" with "fundamental" (see Ney and Phillips 2013, section 3), otherwise (O3) would become self-contradictory. But even so, there is a certain tension between (O3) and the motivation behind the primitive ontology approach. If, as mentioned in section 1, the primitive ontology is introduced in order to account for the existence of macrophysical objects, then (O3) implies that the quantum state by itself is already up to this task, ontologically speaking. One must then ask oneself why a primitive ontology should be postulated at all. Indeed, a large part of the opposition to the primitive ontology approach comes in the guise of what is known as "wave function realism", which denies the need to introduce a primitive ontology, on the grounds that the wave function ontologically suffices to account for the macrophysical world we experience (see the essays collected in Ney and Albert 2013 for recent arguments for and against wave function realism). We therefore do not see much motivation for combining the primitive ontology approach with a commitment to (O3), although such a combination might be a conceptual possibility.

On the other hand, the primitive ontology approach is not automatically committed to (O1). In fact, most discussions of such approaches seem to endorse a variant of (O2) by emphasizing their dual structure, in the sense that the primitive ontology *and* the quantum state together constitute the fundamental ontology of the theory (see, e.g., Allori et al. 2008, section 4).

That the notion of primitive ontology does not, by itself, imply any ontological priority in the sense of (O1) is also emphasised by Maudlin (2013, p. 144; he uses the word "primary" instead of "primitive", but the two notions can be treated as equivalent for our purpose):

> The division of the whole Ontology into Primary and (let us say) Secondary has an epistemic cast. This is not a distinction into different *kinds of existence*. It is a distinction that should track which parts of a theory, according to the theory itself, are more directly and unproblematically related to empirical data and which are more remote from empirical data, and hence more speculative.

Consequently, a theory (if it has a clear story to tell about its relation to empirical data) yields a distinction between a primary (primitive) and a secondary (nonprimitive) ontology, but it does not determine the ontological status of the two parts. Section 2 illustrated this by showing that BM can either be understood along the lines of (O1), as is the case for Humeanism about $\psi$, or along the lines of (O2), as in the views that take $\psi$ to refer to an object or to a law in the primitivist sense. (The case of dispositionalism is somewhat delicate: *prima facie*, it seems to be committed to (O1), since an instantiated property ontologically depends on its bearer. However, it would be misguided to equate the primitive/nonprimitive distinction with the substance/property distinction, because the crucial difference in Bohmian dispositionalism is between two kinds of properties of the particles, namely their positions (which are represented by what is sometimes called the *primitive variables* $Q_1, \ldots, Q_N$) and their quantum state (represented by $\psi$, the *nonprimitive variable*). And none of these two properties can claim ontological priority, which is to say that dispositionalism is committed to a variant of (O2).)

Nevertheless, when it comes to GRWm, it might be tempting to try and read off ontological relations of the mathematical structure of the theory. The reason is that there is in GRWm a mathematical dependence between the primitive and the nonprimitive variables which is absent in the BM case: according to (3), the wave-function at a time $t$ completely determines the primitive variable $m(\cdot, t)$. By contrast, in BM the primitive variables $Q_1(t), \ldots, Q_N(t)$ are not determined by $\psi(t)$, but need to be specified in addition to it. Bacciagaluppi (2010, p. 19) draws the following ontological conclusion from this difference:

> While a dualist position ... is perfectly justified in a no-collapse beable theory, where the beable is defined independently of the wave function, it seems less well motivated in the present case, in which the beable is functionally dependent on the wave function. Indeed, the beable appears to be ontologically completely derivative, so that a dualist interpretation collapses into a wave-only interpretation: I consider this wave-only interpretation to be the most natural reading of a mass-density interpretation.

Such an inference from functional (mathematical) to ontological dependence must, however, be regarded with suspicion. In particular, the asymmetry of the functional dependence (the wave-function completely determining the primitive variable, but not *vice versa*) should not be taken to imply that the primitive ontology (the beable) is ontologically derivative. Consider a related case: there is a similar asymmetric functional dependence between state vectors in Hilbert spaces and the corresponding rays, in that each state vector completely determines its ray, but not *vice versa*. Nevertheless, it is generally agreed that physical states are represented by rays rather than vectors, so one should not ascribe ontological priority to the latter.

Furthermore, Bacciagaluppi's preference for option (O3) squares badly with the basic idea of the primitive ontology approach, as noted above. He calls the mass density "a local manifestation of the wave function" (p. 18). But again, if we had a convincing story about how a wave-function on configuration space can give rise to manifestations in physical space (a story which Bacciagaluppi does not provide), there would be no need to postulate a primitive ontology in the first place. Hence, there is a good reason to resist the idea of reading ontological dependencies off of the functional dependencies contained in the mathematics.

Let us conclude this discussion with two additional remarks. First, it has to be admitted that the above-mentioned analogy between Hilbert space rays and the $m$-variable is not perfect. While the additional structure contained in state vectors as opposed to rays can be seen as an ontologically insignificant mathematical surplus, the additional structure contained in $\psi$ as opposed to $m$ has physical significance. For instance, the spin state of a system at a time $t$ is encoded in $\psi(t)$ but not in $m(\cdot, t)$. The weight one gives to this difference will impact how one chooses between (O1) and (O2). Second, the intimate connection between $\psi$ and $m$ expressed in equation (3) shows that (O2) in the context of GRWm cannot be understood as asserting the mutual independence of $\psi$ and $m$. The idea is rather that the two are interdependent in such a way that none of them can claim ontological priority over the other.

5.   *Assessing the options: differences between Bohmian mechanics and GRWm*

Given the common structure of BM and the primitive ontology theories of GRW described in section 1, we can with good reason take for granted that all the options for understanding the quantum state in BM described in section 2 are, in principle, also available for GRWm. The question then is in what respect the assessment of the different options changes when we switch from BM to GRWm.

The commitment to the quantum state as an object in BM is motivated by wave-particle dualism. Thus, considering the double slit experiment, a common way to introduce BM is to say that the particle goes through one of the two slits and that its motion is guided by a wave that goes through both slits and thereby accounts for the particle distribution on the screen manifesting an interference pattern. Of course, this explanation breaks down as soon as one considers more than one particle, since in this case, the wave-function can no longer represent an ordinary wave or field that propagates in three-dimensional space. It is then rather mysterious how a field on configuration space could guide the motion of particles in three-dimensional space or what a multi-field in three-dimensional space could be. These problems persist if we turn to GRWm, because the mysteries are not resolved by replacing the Bohmian particle ontology with the matter density ontology of GRWm. Moreover, the original motivation for this view, stemming from Bohmian wave-particle duality, vanishes. In GRWm, the primitive ontology consists itself in a field in three-dimensional space, and particle-like behaviour is accounted for in terms of a spontaneous concentration of the matter density around certain points in physical space. There thus is no need for a field over and above the matter density field in order to accommodate wave-particle dualism.

A possible way to spell out the causal relation between the quantum state and the Bohmian particles was recently outlined by Egg and Esfeld (2014, sections 3 and 4), building on a proposal by Blondeau and Ghins (2012). Briefly put, the idea is that anything that figures on the right hand side of a differential equation describing the temporal development of a physical magnitude is a cause of that temporal development. The Bohmian guiding equation

(1) then counts as a causal law, with $\psi$ representing the cause of the temporal development of the particle positions. This proposal does not work for GRWm, because equation (3), which defines the relationship between the quantum state and the matter field, does not contain a time derivative. It hence cannot be regarded as a causal law in the sense of Blondeau and Ghins.[4] In short, what motivates the view of $\psi$ as an object in BM does not apply to GRWm: in this theory, the quantum state cannot be an object that guides or pilots the temporal development of the primitive ontology. (We will see below that it is possible to ascribe a causal role to the quantum state even in GRWm, but not in a way that would motivate the view of $\psi$ as an object existing independently of the primitive ontology, exerting a causal influence on it.)

The general principle underlying the Humean view of the quantum state is the same in BM and in GRWm. In both cases, the quantum state supervenes on the distribution of the primitive ontology throughout the whole of space-time (the particle trajectories, the matter field), figuring in the law that achieves the best combination of simplicity and informativeness as regards that distribution. It is a contingent matter of fact that in a Bohmian particle world, that law is deterministic and time-reversal invariant, whereas in a GRWm world, it is stochastic and not time-reversal invariant (see Frigg and Hoefer 2007 for a Humean view of the GRW probabilities). However, as Allori et al. (2008, sections 6-7) show, one could also combine a primitive ontology of particle trajectories with a stochastic law (as in fact done in Nelson's quantum mechanics, see Nelson 1966, 1985 and Goldstein 1987) and a primitive ontology of a continuous matter density field with a deterministic law. As regards a GRW-type law, Dowker and Herbauts (2005) set out a model according to which the supervenience basis for the quantum state does not have to include the whole distribution of the primitive ontology throughout space-time, but a large enough history of the development of the elements of the primitive ontology is sufficient to fix the quantum state, at least for all practical purposes.

There is thus a considerable flexibility in implementing the thesis of Humean supervenience of the quantum state on the primitive ontology. Nevertheless, the attractiveness of Humeanism differs from one theory to another. In particular, the discussion in the previous section gives us a reason to be less attracted to Humeanism in GRWm than in BM. While Humeanism holds that the quantum state supervenes on the complete history of the primitive ontology, the mathematical structure of GRWm seems to imply just the opposite, namely that the matter density (represented by the primitive variable *m*) supervenes on the quantum state (represented by the nonprimitive variable $\psi$). This is not the case in BM, where different particle configurations are compatible with one and the same quantum state. Now we saw in the previous section that a certain revisionary attitude towards what seems to be implied by the mathematics of GRWm is unavoidable if one takes the idea of a primitive ontology seriously, but a metaphysics that completely reverses what is suggested by the mathematical structure of the theory itself should only be adopted if no other (less revisionary) option is available. And the non-Humean options clearly are less revisionary in this sense, because they all grant some ontological autonomy to the quantum state.

---

[4] There are, of course, other ways to understand causality in physics, but this debate is beyond the scope of the present paper. All we need to claim here is that the Blondeau/Ghins proposal is a reasonable way to spell out the causal character of BM.

On primitivism about laws, in a Bohmian particle world, there simply is a fact instantiated in that world that a deterministic and time-reversal invariant law holds in it, whereas in a GRWm world, there simply is a fact instantiated in that world that a stochastic law holds in it, which is not time-reversal invariant. As in the Humean account, this difference concerns the content of the primitive ontology and the law, but not the relationship between both of them. In that respect, the assessment of the primitivist-nomological approach to $\psi$ does not change when we turn from BM to GRWm.

However, as we have now repeatedly mentioned, there is a difference between BM and GRWm concerning the functional dependence of the primitive on the nonprimitive variables, and this has an impact on the evaluation of primitivism as well. A well-known objection against a nomological understanding (in the primitivist sense) of $\psi$ is that nomological entities should not change in time, whereas $\psi$ in general does so (see Belot 2012, pp. 75-76). In BM, there is a possible reply to this worry, because the wave-function of the universe might be stationary, although the Bohmian particles are moving. (The fact that the Wheeler-DeWitt equation of quantum gravity is time-independent shows that this proposal has some physical motivation and is not just a metaphysical speculation; see Dürr, Goldstein and Zanghì 2013, chapters 11-12.) By contrast, this road is blocked in GRWm, due to the functional dependence of $m(\cdot, t)$ on $\psi(t)$. Within this framework, a stationary $\psi$ would imply a stationary $m$, hence nothing would move in such a universe. (For a GRW-type approach to quantum gravity abandoning the idea of fundamental time-independence, see Okon and Sudarsky 2014, section 4). Therefore, the move from BM to GRWm forces the primitivist to bite the bullet and to admit that the fundamental laws of nature change in time.

As an intermediate summary, we may note that all the three options discussed so far (the $\psi$-as-object view, Humeanism and primitivism) considerably lose attractiveness when transferred from BM to GRWm. In the remainder of this section, we argue that the same is not true for dispositionalism. Let us prepare the ground for this argument by giving a more detailed account of dispositionalism in BM.

On this view, the quantum state as represented by $\psi$ is a property of the whole particle configuration for which it is essential to exercise a certain causal role, namely to determine the velocity of the particles at any time $t$ given the particles' position at $t$ (see Esfeld et al. 2014, sections 4-5). This property hence is a disposition whose manifestation is the velocity of the particles. This disposition does not depend on any external triggering conditions: since it is a property of the whole particle configuration, there is nothing external that could act as a stimulus for its manifestation. Furthermore, it is a sure fire disposition, given that the Bohmian guidance equation is a deterministic law. If the universal wave-function is itself subject to a temporal development (as in the Schrödinger equation), this disposition develops in time; the Schrödinger equation can then itself be regarded as a causal law, describing the cause of that temporal development (see Blondeau and Ghins 2012, p. 386). If the universal wave-function is stationary (as in the Wheeler-DeWitt equation), this disposition does not develop in time, but only the particle configuration does. In brief, on dispositionalism in BM, the quantum state is conceived in analogy to dynamical properties such as mass and charge in classical mechanics: as the distribution of mass and charge in the universe determines the temporal development of the particles' velocities via force laws, so the universal wave-function determines the temporal development of the particles' positions via the guiding equation. The differences are the following three: (a) mass and charge are dispositional properties of the particles taken individually, whereas the quantum state is a dispositional property of the particle configuration as a whole; (b) the manifestation of mass and charge consists in the acceleration of the particles, whereas the manifestation of the quantum state consists in their velocities; (c) mass and charge are time-independent properties, whereas the quantum state is, as things stand, time-dependent.

When applying this view to GRWm, one can retain the main idea of dispositionalism, namely that the job of the quantum state is to exercise a certain causal role in the temporal development of the primitive ontology, although the GRWm law (3) cannot be considered as a causal law like the Bohmian guiding equation (1) (cf. the remark above). Instead, the causal role of the quantum state derives from a causal reading of the law describing the temporal development of the quantum state itself, analogous to the causal reading of the Schrödinger equation in BM mentioned in the previous paragraph. Due to the functional dependence of $m$ on $\psi$ expressed in (3), the quantum state, by virtue of having a causal role in its own temporal development, also has such a role in the temporal development of the matter density. However, the main difference with respect to the situation in BM is that the quantum state in GRWm cannot be a sure fire disposition, since the GRW law is a stochastic equation, not a deterministic one. (To be precise, the GRW law combines the deterministic Schrödinger equation (2) with the stochastic evolution (4) occurring at random times. Since this is rather inelegant, one should remember that the GRW theory is only a simplified approximation to the more realistic CSL theory (see section 3), which describes the complete time evolution of the quantum state by means of a single stochastic differential equation; see Bassi and Ghirardi 2003, equation (8.45).) Accordingly, the quantum state has to be conceived as a probabilistic disposition, which in today's literature is called "propensity". Again, this propensity is a property of the matter density field as a whole, and it manifests itself spontaneously, there being nothing external to the matter density field as a whole that could trigger the manifestation of this propensity.

Hence, the role of the quantum state conceived as a propensity in GRWm is to ground the GRW probabilities. In other words, this conception of the quantum state results from applying the propensity theory of probabilities to GRWm (see Dorato and Esfeld 2010 for propensities in GRW and Suárez 2007 for propensities in QM in general). Since a propensity can be conceived as a special type of a disposition, or the other way round a disposition as a special type of a propensity (namely a sure fire propensity), the transition from sure-fire dispositions to propensities does not affect the attractiveness of dispositionalism when moving from BM to GRWm.

One might object to this claim that propensities are indeed more problematic than sure-fire dispositions, because there is no straightforward connection between propensities and probabilities (see Suárez 2014 for a discussion and a possible solution of the problem). In response, let us recall that BM needs to account for the irreducibly probabilistic predictions of QM as well, and this obviously cannot be achieved by (deterministic) dispositions alone. What needs to be added to connect BM dispositions to quantum probabilities is a certain hypothesis about the initial particle configuration, namely the quantum equilibrium hypothesis (see Dürr, Goldstein and Zanghì 2013, ch. 2). Arguably, the commitment to propensities as the basis of probabilities in GRWm is at least as acceptable as the commitment to sure-fire dispositions *plus* quantum equilibrium required in Bohmian dispositionalism.

Therefore, given the problems arising for the other three options ($\psi$ as object, Humeanism and primitivism) when turning from BM to GRWm, the relative attractiveness of dispositionalism increases in the wake of this transition. And since there are already good reasons to prefer dispositionalism in the context of BM (see Esfeld et al. 2014, section 5), we conclude that dispositionalism is clearly the most attractive option for conceiving the quantum state in GRWm.

*Acknowledgments*: Earlier versions of this paper were presented at the spring meeting of the German Physical Society (Berlin, March 2014), the CROSS workshop on the quantum state (Lausanne, May 2014) and the II PERSP workshop on space-time and the wavefunction (Barcelona, May 2014). We thank the participants of these events, especially Wayne Myrvold, Nino Zanghì, Guido Bacciagaluppi and Jeremy Butterfield, for many helpful remarks.